\documentclass[]{spie}  %>>> use for US letter paper
\usepackage{amsmath,amsfonts,amssymb}
\usepackage[]{graphicx}
\usepackage[colorlinks=true, allcolors=blue]{hyperref}
\usepackage{orcidlink}

\title{CCAT: A status update on the EoR-Spec instrument module for Prime-Cam
}

\author[a]{Rodrigo Freundt \orcidlink{0000-0002-8169-538X}}
\author[b]{Yaqiong Li}
\author[c]{Doug Henke}

\author[d]{Jason Austermann}
\author[e]{James R. Burgoyne}
\author[c,e,f]{Scott Chapman}
\author[g]{Steve K. Choi \orcidlink{0000-0002-9113-7058}}
\author[b]{Cody J. Duell}

\author[b]{Zach Huber}
\author[b,a]{Michael Niemack}
\author[h]{Thomas Nikola}

\author[b]{Lawrence Lin}
\author[i]{Dominik A. Riechers}
\author[a]{Gordon Stacey}
\author[d]{Anna K. Vaskuri}
\author[b,j]{Eve M. Vavagiakis}
\author[d]{Jordan Wheeler}
\author[k]{Bugao Zou}

\author{the CCAT collaboration}

\affil[a]{Department of Astronomy, Cornell University, Ithaca, NY 14853, USA.}
\affil[b]{Department of Physics, Cornell University, Ithaca, NY 14853, USA.}
\affil[c]{NRC Herzberg Astronomy and Astrophysics, 5071 West Saanich Rd, Victoria, BC, V9E 2E7, Canada.}
\affil[d]{Quantum Sensors Division, National Institute of Standards and Technology, Boulder, CO 80305, USA.}
\affil[e]{Department of Physics and Astronomy, University of British Columbia, Vancouver, V6T 1Z1, Canada.}
\affil[f]{Department of Physics and Atmospheric Science, Dalhousie University, Halifax, NS, B3H 4R2, Canada.}

\affil[g]{Department of Physics and Astronomy, University of California, Riverside, CA 92521, USA}
\affil[h]{Cornell Center for Astrophysics and Planetary Sciences, Cornell University, Ithaca, NY 14853, USA.}
\affil[i]{Institut f\"ur Astrophysik, Universität zu K\"oln, Z\"ulpicher Strasse 77, 50937 K\"oln, Germany.}
\affil[j]{Department of Physics, Duke University, Durham, NC 27708, USA.}
\affil[k]{Department of Applied and Engineering Physics, Cornell University, Ithaca, NY 14853, USA.}

\authorinfo{Further author information: (Send correspondence to R.F.)\\R.F.: E-mail: rgf57@cornell.edu}

\begin{document} 
\maketitle

\begin{abstract}
The Epoch of Reionization Spectrometer (EoR-Spec) is an upcoming Line Intensity Mapping (LIM) instrument designed to study the evolution of the early universe (z = 3.5 to 8) by probing the redshifted [CII] 158 $\mu$m fine-structure line from aggregates of galaxies. The [CII] emission is an excellent tracer of star formation since it is the dominant cooling line from neutral gas heated by OB star light and thus can be used to probe the reionization of the early Universe due to star formation. EoR-Spec will be deployed on Prime-Cam, a modular direct-detection receiver for the 6-meter Fred Young Submillimeter Telescope (FYST), currently under construction by CPI Vertex Antennentechnik GmbH and to be installed near the summit of Cerro Chajnantor in the Atacama Desert. This instrument features an image plane populated with more than 6500 Microwave Kinetic Inductance Detectors (MKIDs) that are illuminated by a 4-lens optical design with a cryogenic, scanning Fabry-Perot Interferometer (FPI) at the pupil of the optical system. The FPI is designed to provide a spectral resolving power of $R\sim100$ over the full spectral range of 210--420 GHz. EoR-Spec will tomographically survey the E-COSMOS and E-CDFS fields with a depth of about 4000 hours over a 5 year period. Here we give an update on EoR-Spec’s final mechanical/optical design and the current status of fabrication, characterization and testing towards first light in 2026.

\end{abstract}

% Include a list of keywords after the abstract 

\section{INTRODUCTION}
\label{sec:intro}  % \label{} allows reference to this section

An important outcome of the formation of the first stars and galaxies in the early universe is the so-called Epoch of Reionization (EoR), a cosmic time that has been technically difficult to probe due to the intrinsically faint, redshifted emission from these first luminous objects. During the EoR, the primordial hydrogen left after recombination ($z\sim1100$) underwent a gradual changeover from neutral to fully ionized as the strong radiation fields produced by these first stars and galaxies heated the interstellar and intergalactic mediums. There is solid observational evidence from the absorption spectra of distant quasars and from measurements of the integrated optical depth to reionization in CMB experiments that this process was mostly completed at around redshift $z\sim6$\cite{robertsonCOSMICREIONIZATIONEARLY2015a}. Cosmological simulations and targeted observations of high-redshift star forming galaxies suggest that the reionization of the universe was neither homogeneous nor instantaneous, but rather a process that evolved in patches, starting at $z\sim15\text{--}20$, and whose evolution was possibly correlated with the star formation history of the universe\cite{madauCosmicStarFormationHistory2014}. We are thus interested in the details of the reionization history, such as the characteristic scale of the reionization bubbles, if any, and the statistics of the population of sources that drove this process. From a cosmological viewpoint, building a picture of the EoR by tracking the ionization fraction over cosmic time is of paramount interest, as it might encode information about the expansion of the universe, which can help break tensions in cosmology, as well as shed light on the nature of dark matter.

While galaxy surveys have probed deeper and deeper into the high redshift universe\cite{atekMostPhotonsThat2024}, it is still technically infeasible to survey the whole population of sources over large cosmic volumes. An alternative technique called Line Intensity Mapping (LIM)\cite{bernalLineintensityMappingTheory2022} has been proposed to overcome this difficulty. LIM maps atomic and molecular lines in aggregate without resolving individual line emitters. By using the spectroscopic redshift of specific target lines, LIM can tomographically build three-dimensional intensity maps (spectral cubes) of large cosmological volumes such as the EoR and cosmic noon. In the past decade, LIM has seen wider adoption in the experimental astrophysics and cosmology community with fielded experiments such as CHIME, COMAP, and TIME, among many others\cite{bernalLineintensityMappingTheory2022}.

The Epoch of Reionization Spectrometer (EoR-Spec)\cite{nikolaCCATprimeEpochReionization2022} is an imaging spectrometer designed to obtain tomographic maps of the late EoR ($z \sim 3.5 \text{--} 8.0$) through LIM of fluctuations in the aggregate clustering signal encoded in the 158 micron fine-structure transition line from ionized carbon. This emission originates primarily\footnote{In star-forming regions, the dominant contribution ($\sim70\%$) to the observed [CII] emission is from PDRs. However, a very significant fraction of the observed [CII] emission could come from ionized gas when observing this line on scales larger than individual galaxies).} in Photo Dissociation Regions (PDRs) at the surface of dense molecular clouds, and is the dominant cooling mechanism for star formation in these regions\cite{stacey158MmIi2010}. [CII] has been proven to be an excellent tracer of star formation, both in the local and early universe, where it is redshifted to sub-mm wavelengths and thus accessible from the ground with facilities like the Fred Young Submillimeter Telescope (FYST)\cite{collaborationCCATprimeCollaborationScience2022}, which our collaboration is building near the top of Cerro Chajnantor at an elevation of 5600-m in the Atacama desert. Prime-Cam\cite{choiSensitivityPrimeCamInstrument2020,Vavagiakis_2018} is a cryogenic, direct-detection receiver for FYST and populates the focal plane with up to seven modular instruments. The EoR-Spec instrument modules will be dedicated to carry out the CCAT Deep Spectroscopic Survey (DSS), a 4000-hour LIM survey of E-COSMOS and E-CDFS fields. EoR-Specs' angular (sub arc-minute) and spectral resolution ($R\sim100$) are matched to the expected clustering signal. Thus, a significant detection of [CII] at $z\lessapprox6$ is expected\cite{clarkeCIILuminosityModels2024,karoumpisCIILineIntensity2022}, as well as important limits at higher redshifts\cite{chungForecastingIILineintensity2020}.

In this paper, we give an update on the final design of EoR-Spec. In section \ref{sec:Instru}, we describe the instrument and present changes made to the optomechanical design, while in section \ref{sec:stray-ligth} we comment on potential optical systematics. Finally, in sections \ref{sec:detarray}, we report on recent progress on the fabrication and testing of the detector arrays for this instrument. 

\section{INSTRUMENT DESIGN}
\label{sec:Instru}

EoR-Spec features three large-format arrays of Microwave Kinetic Inductance Detectors (MKIDs), with a total of 6528 non-polarized pixels divided into two bands. Given that we expect a brighter signal at the lowest redshift range, we have weighted the number of detector such that two arrays are sensitive to the 210--315 GHz range or Low Frequency (LF) band, while the third sub-array is sensitive to the 315--420 GHz frequency range or High-Frequency (HF) band. All three arrays will be lumped-element aluminum MKIDs that are front-side illuminated with an aluminum feedhorn array that efficiently couples radiation into the detector. A series of high-purity silicon lenses reimage the telescope's focal plane into the aperture plane of the feedhorns. 

The element that gives the spectral response to the camera design described is the scanning Fabry-Perot Interferometer, which is strategically located at the 4 Kelvin stage and aligned with the pupil of the optics, where collimation ($F/\#>100$) is maximized. A FPI consists of two identical, highly reflective mirrors that form a resonant cavity, sometimes also called the etalon, whose resonant frequency (maximum transmission) can be tuned to a certain wavelength range by increasing the spacing between the mirrors systematically while maintaining high parallelism with respect to each other. For EoR-Spec, we are fielding a novel etalon design with Silicon-Substrate-Based (SSB) mirrors that feature etched metamaterial Anti Reflection Coating (ARC) on one side and a evaporated metal mesh on the other surface, as described in \citenum{zouCCATprimeDesignCharacterization2022} and \citenum{zouSPIE2024}. These SSB mirrors form a resonant cavity that essentially behaves as a variable filter, with a spectral response that also depends on the angle of incidence. We operate the SSB etalon in second and third interference orders, corresponding to the LF and HF bands, respectively.

\begin{figure} [ht]
    \begin{center}
    \begin{tabular}{c} 
        \includegraphics[width=0.9\textwidth]{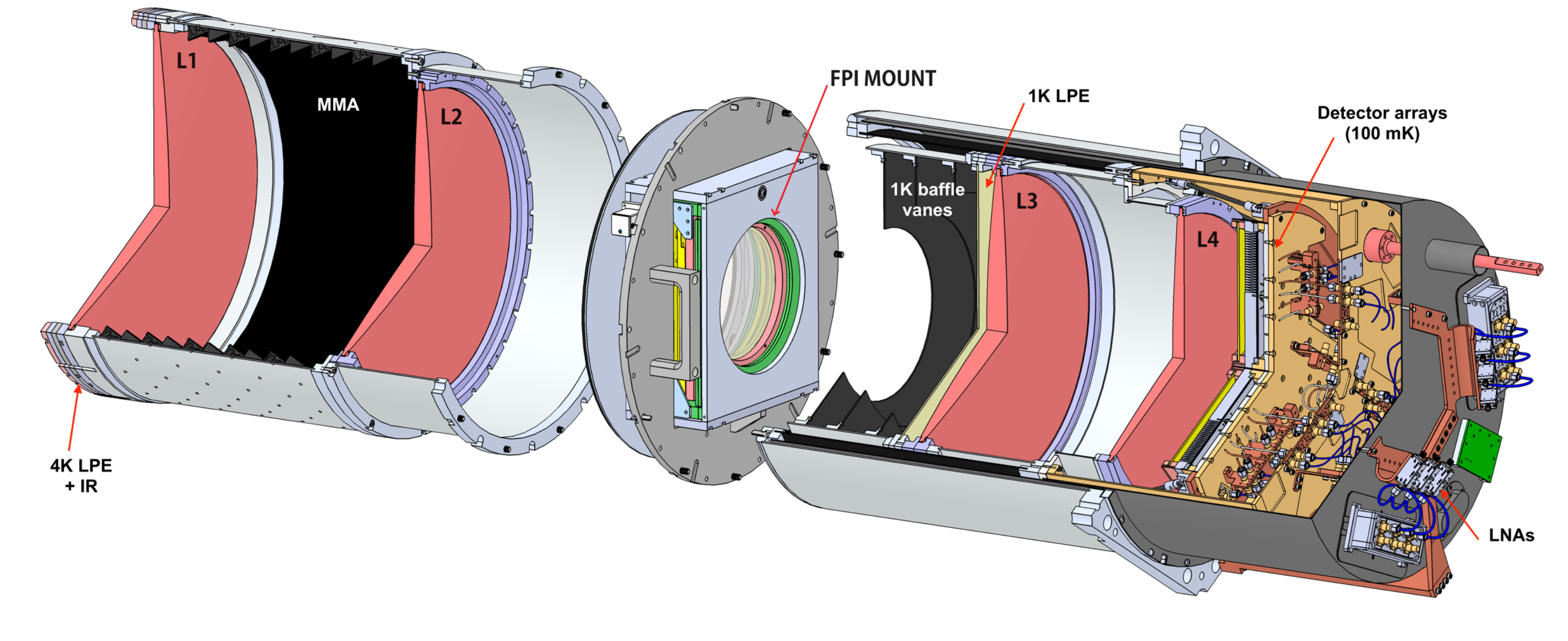}
    \end{tabular}
	\end{center}
    \caption{\label{fig:eor-spec-im} A semi-exploted view of the Computer-Aided Design (CAD) of the EoR-Spec instrument module. The light beam enters from the left and propagates throughout the 4-element (L1-L4) refractive optics design, painted in red, as well as Low-Pass Edge (LPE) filters at 4K and 1K. L1 and L2 are located at the 4 K stage, as well as the FPI (middle section). L3 and L4 are located at 1 K. The 1 K temperature stage also features a 3-vane baffle design for stray light mitigation, which also works as a radiation shield. The three detector arrays are located behind L4. The area between L1 and L2 is covered with Microwave Metamaterial Absorber (MMA) tiles\cite{Zhu_2021} for further stray light mitigation. From end to end, the length of the instrument module is about 1.2 meters.}
\end{figure}

The mechanical design of the instrument module (Figure \ref{fig:eor-spec-im}) has evolved from Prime-Cam's 280 GHz instrument module design \cite{vavagiakisCCATprimeDesignModCam2022} and the Simons Observatory's Large Aperture Telescope Receiver optics tubes\cite{Zhu_2021}. Hence, it inherits the same cryogenic design that consists of rigid aluminum cylinders that mount critical elements along the optical axis (such as lenses, filters and the cold stop), as well as the use of carbon fiber structures to thermally isolate stages at different cryogenic temperatures. EoR-Spec leverages on this general camera designs but adds a fourth silicon lens to the optical train and the FPI at the cold stop. 

Due to the lower cooling power available at 1 K, the cold Lyot stop (and thus the FPI) was relocated to the 4 K temperature stage with minimal additional radiation loading on the detector array. This also makes the FPI easier to mount to the instrument module; by adding an intermediate section to the 4 K tube, the FPI can be easily accessed for troubleshooting during commissioning.

\subsection{Cryogenic Optomechanical Design}

The optical design for EoR-Spec originally required the use of biconic lenses \cite{huberCCATprimeOpticalDesign2022} to meet the specification of the instrument, including high collimation at the Lyot stop ($F/\#>100$) while maintaining high image quality (Strehl ratio above 0.8 over the full $1.3^\circ$ field of view). After many iterations, the final design\cite{huberSPIE2024} now incorporates only aspheric lenses without sacrificing optical performance. Aspheric lenses are easier and cheaper to fabricate than biconic lenses and do not require clocking alignment. Figure \ref{fig:eor-spec-im} shows the four silicon lenses colored red. Not shown on the image are the metamaterial ARCs that have to be machined on the surfaces of the lenses to avoid ghosting (Section \ref{sec:stray-ligth}). We label these lenses with their numbered position from sky to detectors (L1-L4). 

\begin{figure} [ht]
   \begin{center}
   \begin{tabular}{c} %% tabular useful for creating an array of images
   \resizebox{.9\textwidth}{!}{
        \includegraphics[height=3cm]{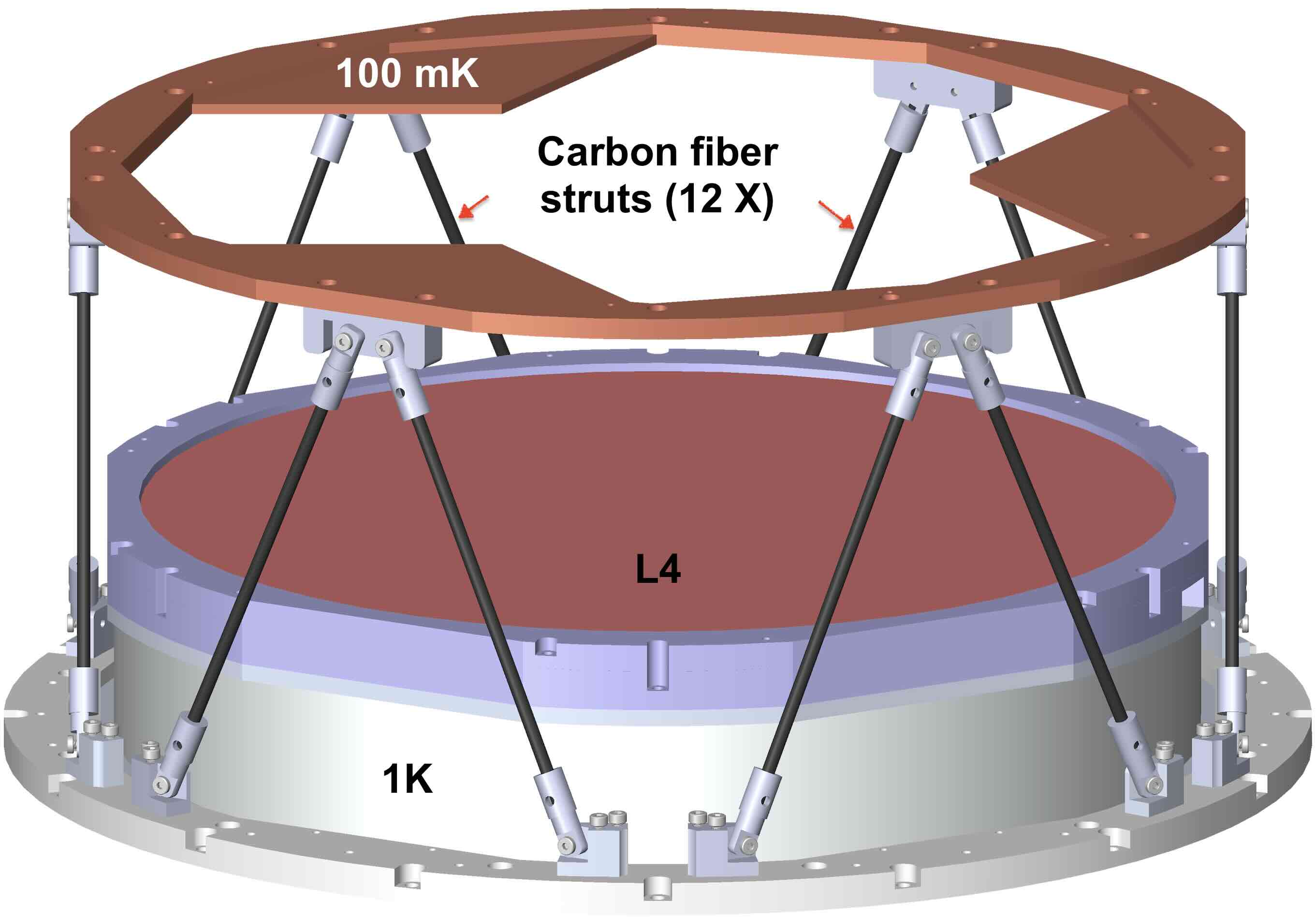}%
        \quad
        \includegraphics[height=3cm]{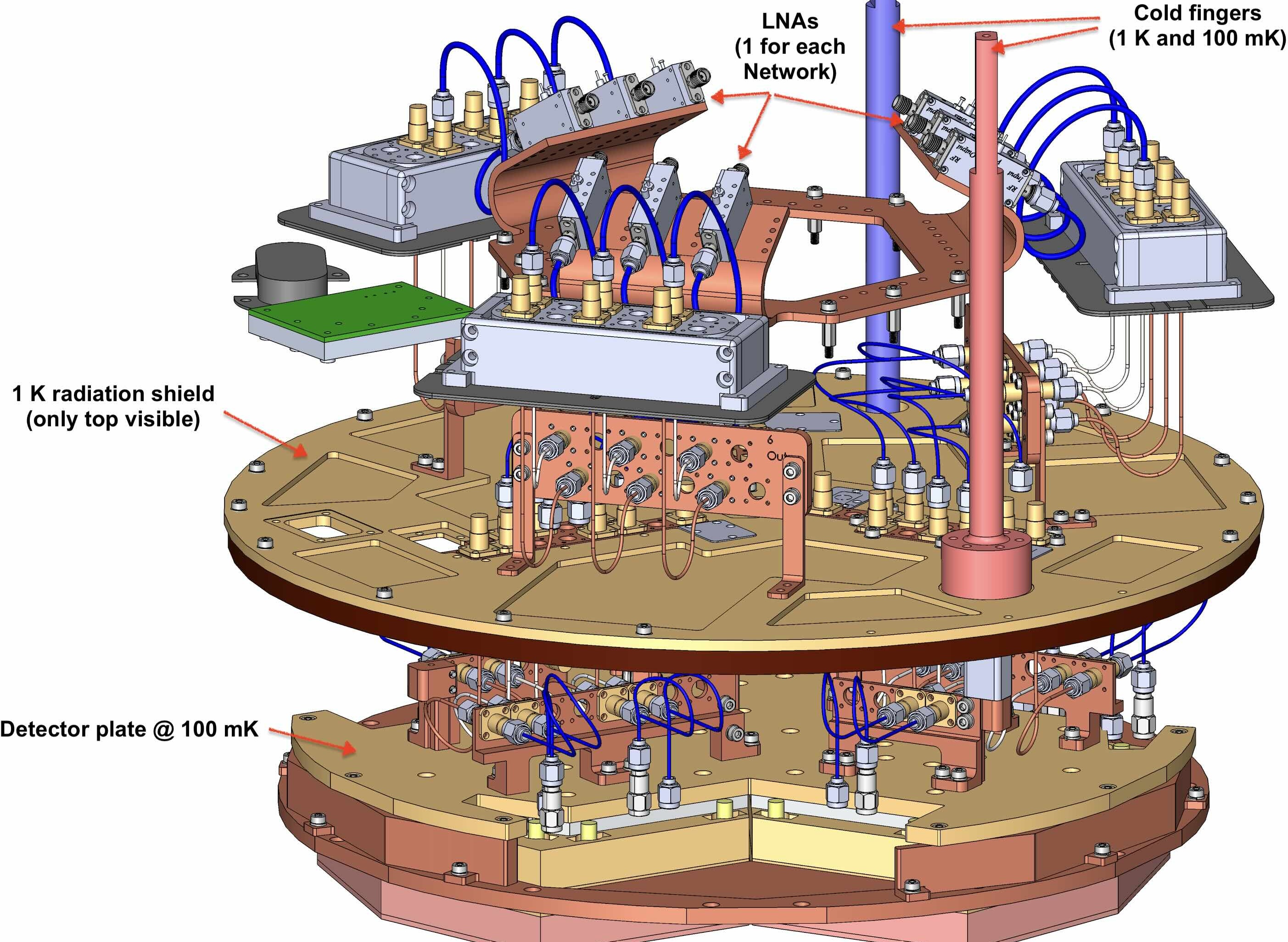}%
    }
   \end{tabular}
   \end{center}
   \caption[example] {\label{fig:optomechanical-design} \textbf{Left:} Detailed view of the 1 K-to-0.1 K interface. Because of the short distance that need to be maintained between L4 and the image plane, the lens cell for L4 was designed such that it fits inside the carbon fiber truss. \textbf{Right:} Close-up view of detector arrays and readout hardware.}
\end{figure}

In this final optical design, the distance between the final lens in the module and the detector focal plane is significantly shorter in EoR-Spec than in Prime-Cam's other instrument modules, making the design of a shorter carbon fiber truss\cite{vavagiakisCCATprimeDesignModCam2022} impossible without increasing thermal conduction from 1 K to 100 mK. One solution was to incorporate L4 to the detector array plate, but this would have increased the heat capacity at the 100 mK stage, where cooling power is limited. Instead, we decided to keep the diameter of L4 small enough to fit inside the carbon fiber truss (Figure \ref{fig:optomechanical-design}), which complicates the assembly procedure but keeps the same truss length and thermal leakage as in Prime-Cam's other modules. Other mechanical constraints set by carbon fiber tube diameter required us to keep the physical diameter of L3 no larger 336 mm.

\subsection{Readout, Thermometry and FPI Control}

MKIDs allow for frequency-multiplexed readout schemes, where a large number of detectors with unique resonant frequencies can be excited and read out simultaneously with a single transmission line. The resonators are designed to have high quality factors and adequate spacing to avoid collisions in the working bandwidth. 

The same cold readout design as the 280 GHz module\cite{vavagiakisCCATprimeDesignModCam2022} has been used for EoR-Spec but has been scaled down to the number of readout channels required. The LF detector arrays need three networks each (two coaxial cables per network) to read out 1728 MKIDs, while the HF array is planned to double the amount of networks to read out a total of 3072 detectors. Low Noise Amplifiers (LNA)\footnote{Designed by Arizona State University.} cooled to 4 K are required at the exit port of each network. A bundle of flexible coaxial cables are routed from the LNAs (depicted also in Figure \ref{fig:optomechanical-design}) to Prime-Cam's readout harness where the coaxial cables transition into flexible striplines that are heat sunk at 4 K and 40 K, to finally be digitized and processed by the warm readout electronics inside the instrument cabin\cite{sinclairCCATprimeRFSoCBased2022}.

Thermometers will be located strategically across the instrument module. Thermometry cables are routed in between the 4 K shell and the carbon fiber shell, as they need to be connected to the readout harness from the back of the module. Triaxial and twisted-pair cables are also required for the capacitive sensors and cryogenic stepper motors, respectively, which are used to scan the FPI during the survey. 

\section{Stray Light Analysis}
\label{sec:stray-ligth}

\begin{figure} [ht]
   \begin{center}
   \begin{tabular}{c} %% tabular useful for creating an array of images
   \resizebox{.9\textwidth}{!}{
        \includegraphics[height=2.5cm]{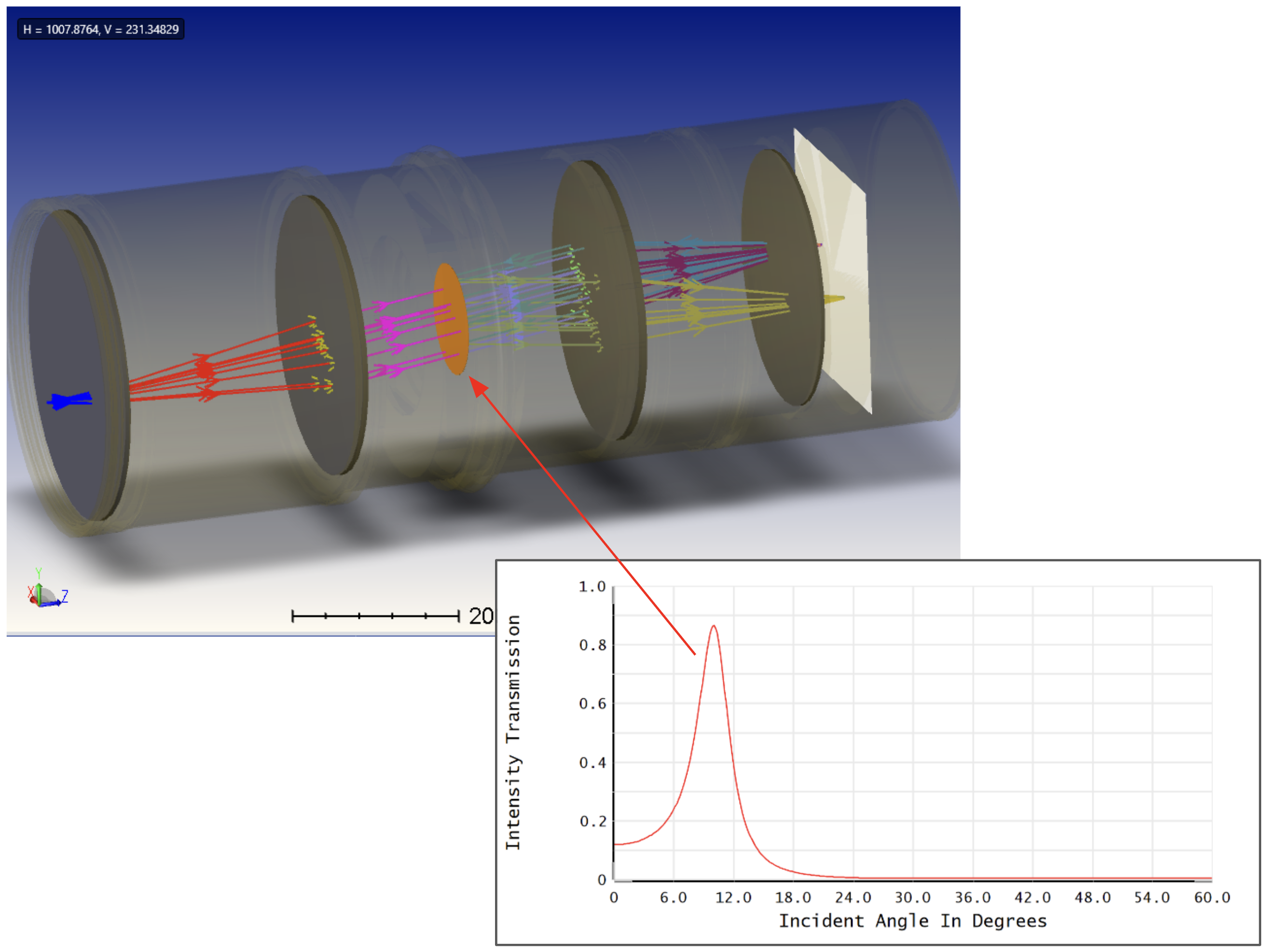}%
        \includegraphics[height=3cm]{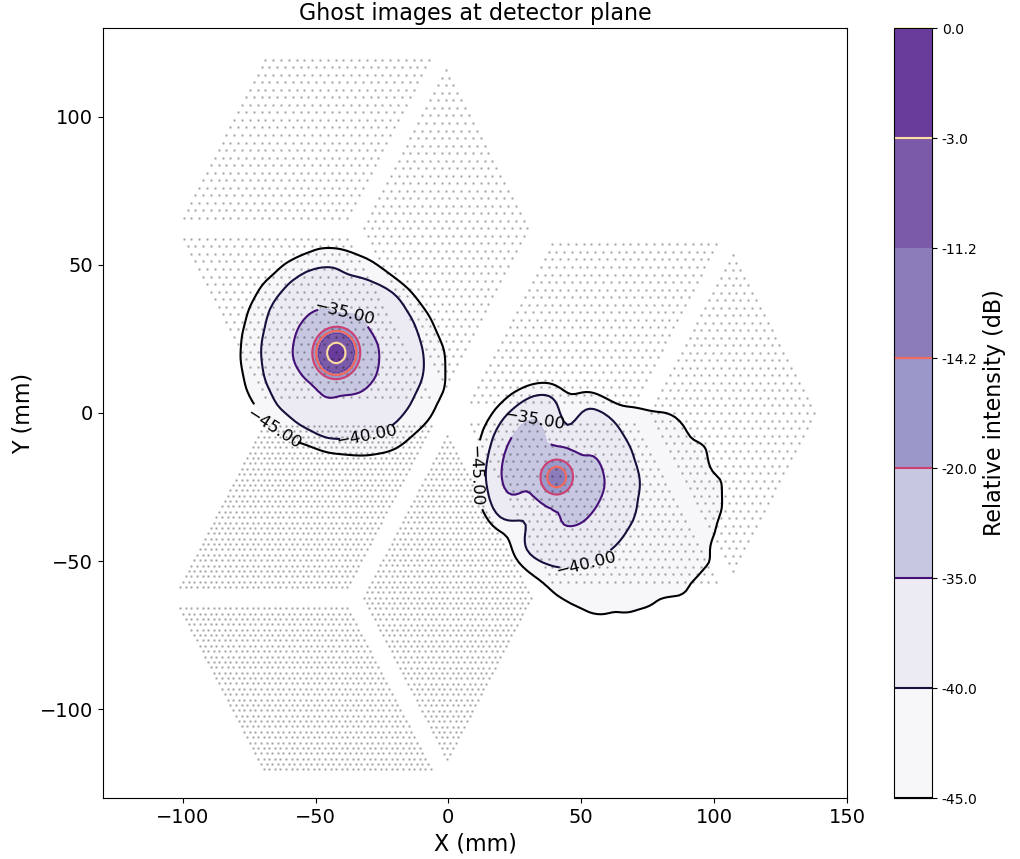}%
    }
   \end{tabular}
   \end{center}
   \caption[example] {\label{fig:stray-light} Stray light analysis. \textbf{Left:} Non-sequential ray trace of an off-axis field point. The ray trace was filtered so that it shows only unwanted reflections from the image plane that interact with the etalon. An uniform scattering profile was assigned to the image plane for simplicity, while the designed SSB mirror transmission profile was assigned to the dummy surface (orange) at the Lyot stop. \textbf{Right:} Spurious images formed at the image plane from unwanted reflections off optical and mechanical components. Normalized and smoothed with the Point Spread Function (PSF). Point-like ghost image on the right side peaks at -14 dB below the primary image, while diffuse ghost images (from reflections off silicon lenses) are at the -35 dB level.}
\end{figure}

Given the major changes (described in Section \ref{sec:Instru}) from the general camera design, we decided to carry out a stray light analysis similar to the ones described in \citenum{fowlerOpticalDesignAtacama2007,gallardoStudiesSystematicUncertainties2018,gudmundssonSimonsObservatoryModeling2021}, but focused on the effects of unwanted reflection from the FPI structures, including the etalon. Figure \ref{fig:stray-light} shows a non-sequential light path in \texttt{ZEMAX Optic Studio}, where light reflected back and forth from the image plane and the FPI mirror form a secondary image. This ghost image is symmetric with respect to the center of the image for any given field angle, and is produced because both the etalon and the feedhorn surface are parallel to each other and the optics is telecentric. \footnote{Introducing a small tilt to the etalon does not avoid intercepting the secondary image. A higher tilt angle would shift the frequency response and reduce the clear aperture.}

\begin{figure} [ht]
   \begin{center}
   \begin{tabular}{c} %% tabular useful for creating an array of images
   \resizebox{.9\textwidth}{!}{
        \includegraphics[height=3cm]{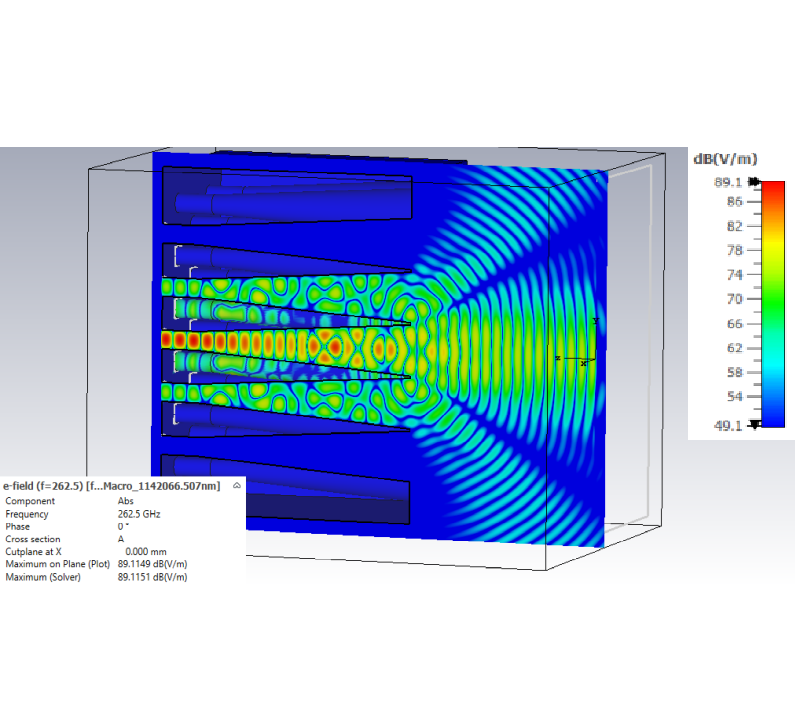}%
        \quad
        \includegraphics[height=3cm]{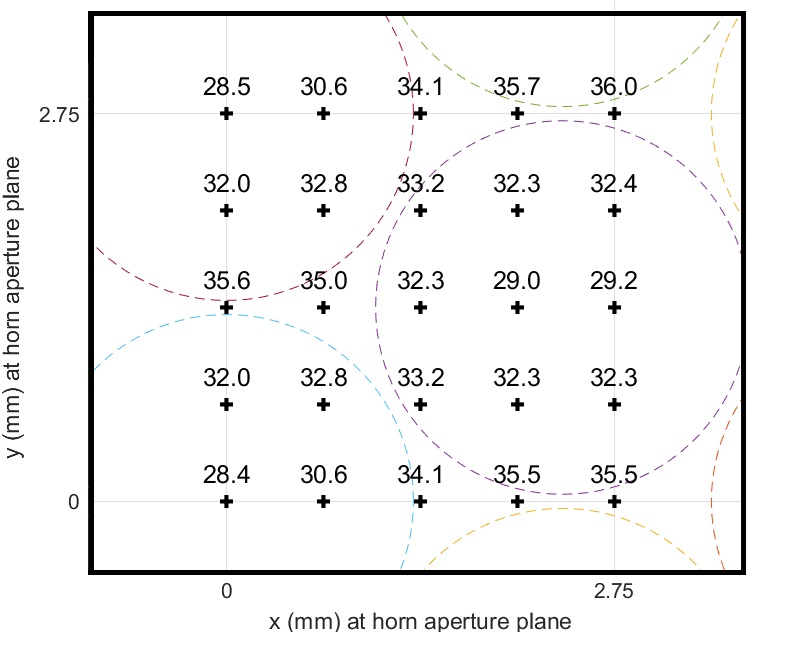}%
    }
   \end{tabular}
   \end{center}
   \caption[example] {\label{fig:reflections-feehorn} Scattering analysis for $f=262.5$ GHz. \textbf{Left:} \texttt{CST} simulation of an on-axis Gaussian beam focused at the aperture plane of the feedhorn array (49 horns), with the beam waist $0.84\times\lambda\times N = 2.015$ mm to approximate a beam with f/\# of 2.1. Figure shows a slice of the scattering distribution function. \textbf{Right}: Scattered power ($\%$) for an on-axis Gaussian beam at different offsets across the array.}
\end{figure}

To determine the level of ghosting, a suite of simulations were carried out in \texttt{CST studio} software to understand scattering off the feedhorn array surface. For this, a on-axis Gaussian beam was scanned across a $7\times7$ feedhorn array section, including the highly-reflective metal areas between the feedhorns.  The waist of the beam was approximated to fit an f/\# of 2.1. The percentage of the total power reflected off this interface and out of the simulation box (i.e., not coupled into any adjacent feedhorn) is shown in Figure \ref{fig:reflections-feehorn} for an array of 25 locations. This reflectivities take into account both specular and non-specular reflections. We combined these results, along with the designed transmission profile of the SSB mirrors, to determine that the spurious image is approximately $-14$ dB below the main image. This is a pessimistic estimate because we only study on-axis reflections (with respect to the feedhorn aperture plane) and we assume that all scattered power reflects specularly rather than considering a proper scattering distribution function. Regardless, the ghosting level is still comparable with the diffraction limit and thus will not affect observations. 

The same suite of simulations was used to confirm the sections of the instrument module that needed to be coated with absorbers or else they could introduce undesired loading onto the detectors if a fraction of those stray-light paths ends up at 300 K (from a time-reverse perspective). We decided to coat the front of the FPI structure with an microwave absorber material, which could be either a metamaterial tiles like in \citenum{Zhu_2021} or cryogenic epoxy loaded with charcoal.  

\section{LOW FREQUENCY DETECTOR ARRAY}
\label{sec:detarray}

We have started fabricating the first LF detector array for EoR-Spec. Two out of three arrays at the image plane are LF arrays, with 1728 pixels each and band centered at 265 GHz. Each hexagonal array is divided into 3 rhombus-shaped groups of pixels that are read out with a single transmission line. Figure \ref{fig:feedhorn-figures} shows the fabricated aluminum box that houses the detector's silicon wafer stack inside. The design of the detector box and the assembly procedure is described in more detail in Ref. \citenum{liCCATprimeDesignEpoch2022}.

\subsection{Feedhorn Array Design and Fabrication}

The feedhorn array is based on previous designs of close-packed, simple conical horns like in \citenum{mccarrickHorncoupledCommerciallyfabricatedAluminum2014} and \citenum{austermannMillimeterWavePolarimetersUsing2018}. The spacing of feedhorns is determined by the pixel pitch of 2.75 mm (1.2 F$\lambda$), optimized from calculations of mapping sensitivity\cite{choiSensitivityPrimeCamInstrument2020}. For an $f/\#$ of 2.1, we have optimized the aperture diameter, flare length and wall thickness for maximum coupling efficiency, spillover efficiency and ease of machining.

\begin{figure} [ht]
   \begin{center}
   \begin{tabular}{c} %% tabular useful for creating an array of images
   \resizebox{.9\textwidth}{!}{
        \includegraphics[height=3cm]{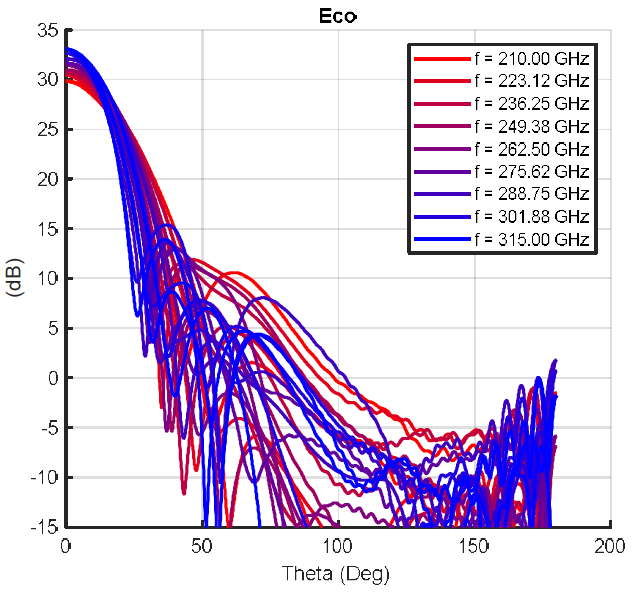}%
        \quad
        \includegraphics[height=3cm]{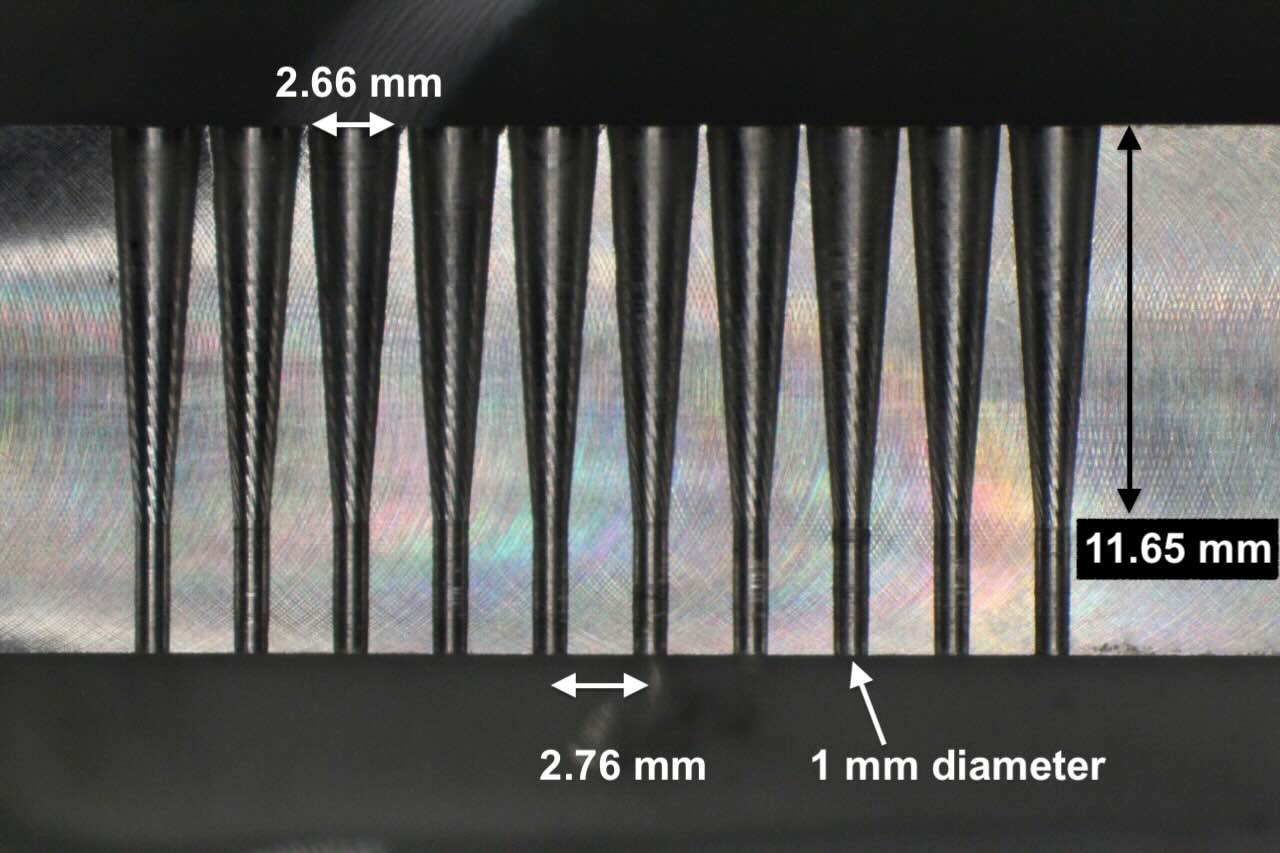}%
        \quad
        \includegraphics[height=3cm]{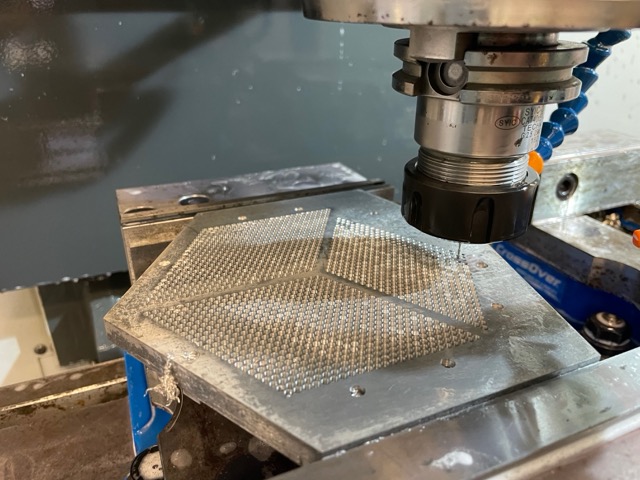}%
    }
   \end{tabular}
   \end{center}
   \caption[example] {\label{fig:feedhorn-figures} LF feedhorn array design and fabrication. \textbf{Left:} Half power bandwidth (HPBW) of the feedhorn design. HPBW ranges from 20 to 35 degrees depending on azimuthal cut and the operating frequency. \textbf{Middle:} Array of 10$\times$10 feedhorn fabricated to test the machining process. The sample was sliced through the first and last rows to reveal the feedhorn profile. \textbf{Right:} Full LF feedhorn array under fabrication. We expect that the nanometer-scale oxidation layers formed on the QC-10 alloy from the coolant used during the milling process will not affect the coupling efficiency at the wavelengths of interest.}
\end{figure}

The feedhorn array was fabricated with a conventional CNC milling approach using a series of custom cutters on a monolithic block of \texttt{ALCOA QC-10} alloy. The material of choice was based on ease of machining and the fact that the material is stress-relieved. The machining procedure involves first the removal of the bulk of the material with a series of rough cutters, and second the use of custom-made taper-ball-end cutters and reamers that drill the final shape of the feedhorn profile as shown in Figure \ref{fig:feedhorn-figures}. To avoid cutter failure, sets of $\sim100$ feedhorns were drilled at a time before exchanging cutters, allowing uniform horn surface roughness and hole roundness throughout the process. This number was determined from the fabrication of prototypes (middle pane in Figure \ref{fig:feedhorn-figures}) that were later sliced to measure the surface with feedhorns with a profilometer.

\subsection{Detector Array Fabrication and Testing}

\begin{figure} [ht]
   \begin{center}
   \begin{tabular}{c} %% tabular useful for creating an array of images
   \resizebox{.9\textwidth}{!}{
        \includegraphics[height=3cm]{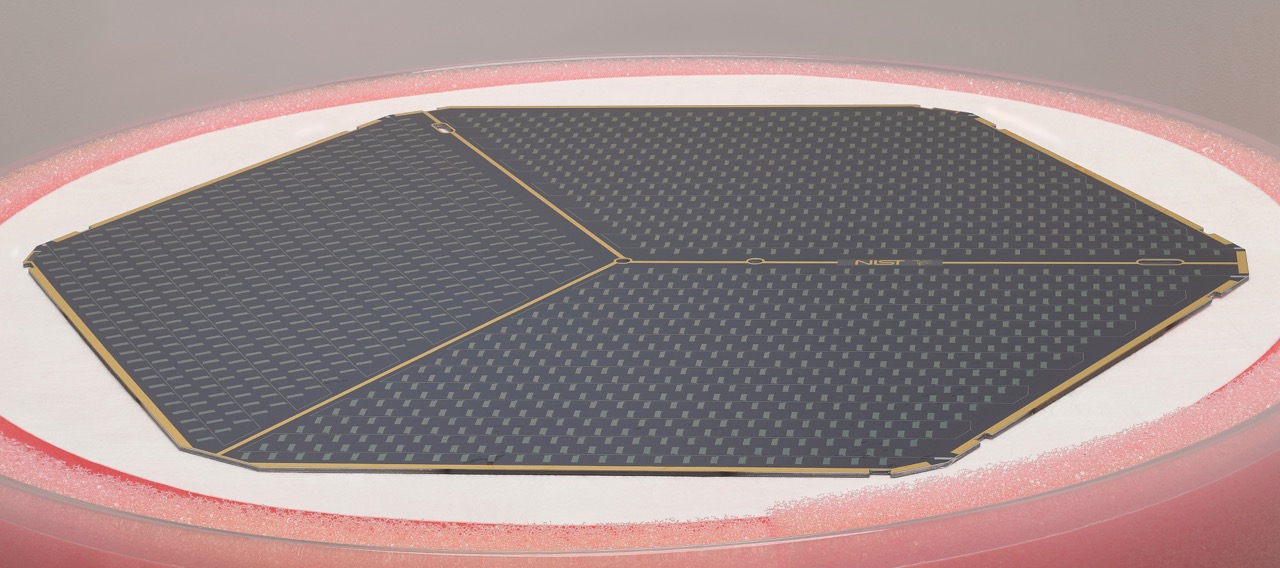}%
        \quad
        \includegraphics[height=3cm]{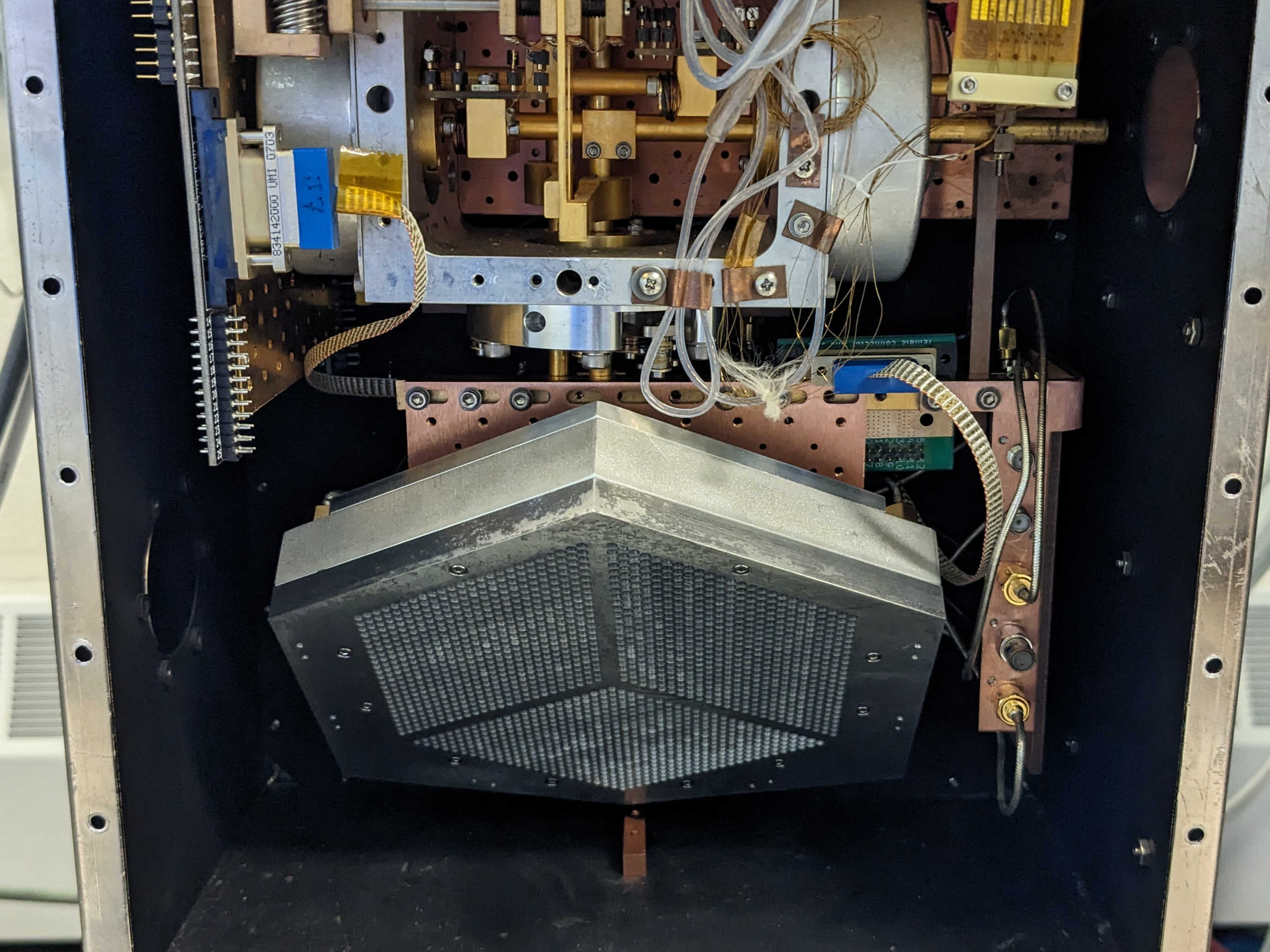}%
    }
   \end{tabular}
   \end{center}
   \caption[example] {\label{fig:NIST-figures} Low Frequency MKID array fabrication and testing. \textbf{Left:} Photograph of detector array wafer, where inter-digitated capacitors and feedlines are visible. \textbf{Right}: Detector box with the silicon wafer stack was mounted inside an ADR for testing.}
\end{figure}

\begin{figure} [ht]
    \begin{center}
    \begin{tabular}{c} 
        \includegraphics[height=6cm]{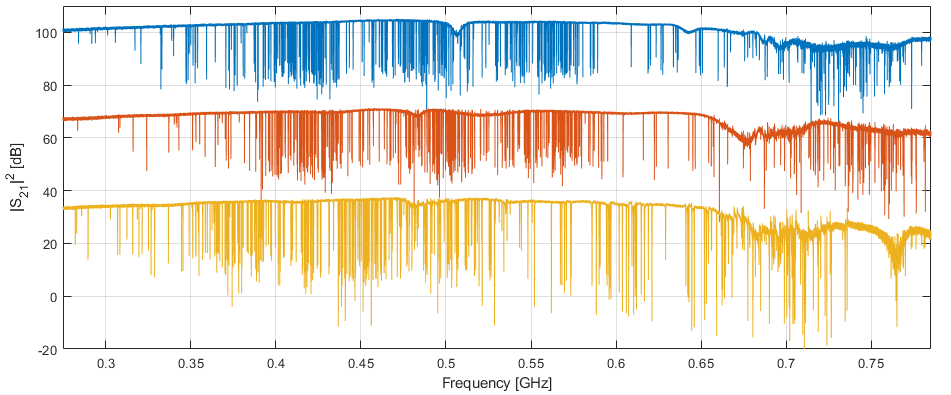}
    \end{tabular}
    \end{center}
    \caption{\label{fig:s21-graph} Preliminary $S_{21}$ for each of the three channels of the LF array. A detector yield of 85\% was accomplish with only the detector wafer and spacer wafer. This yield is expected to improve in the final configuration, where better grounding is achieved.}
\end{figure}

The LF detector array shown in Figure \ref{fig:NIST-figures} has been fabricated by the Quantum Sensors Division at the National Institute of Standards and Technology (NIST). The full silicon wafer stack consists of the detector array, the Waveguide Interface Plate (WIP), a spacer wafer, and a blank wafer, as described in more detail in ref. \citenum{liCCATprimeDesignEpoch2022}. For the first cryomechanical test, we placed only the detector array and the spacer wafer inside a dark box at cryogenic temperatures, and ran a frequency sweep to locate the resonators (Figure \ref{fig:s21-graph}), resulting in a preliminary yield of 85\%. Figure \ref{fig:NIST-figures} also shows our current test setup, where the full wafer stack in placed inside the detector array box (with feedhorns included), which is mounted to the milli Kelvin stage of an Adiabatic Demagnetization Refrigerator (ADR). We read out each of the 3 channels with a single coaxial cable by using a cold RF switch. 

The next steps planned for the LF array characterization is to carry out LED mapping to pinpoint and assign resonators to individual pixels in the focal plane, as well as characterizing the optical performance of the detectors using a cold load at different tone powers\cite{AnnaSPIE2024}. Based on the results, we will evaluate if collisions and cross-talk can be resolved with capacitor trimming to increase the yield of this array.

\section{CONCLUSIONS AND FUTURE WORK}

Mapping [CII] in a continuous redshift range with a background-limited imaging spectrometer like EoR-Spec will enable direct mapping of the large-scale distribution of the sources of reionization in the early universe. The final optomechanical design for EoR-Spec has been presented, where a careful analysis of critical components was carried out to maximize instrument performance. Technical drawings for the 4 K and 1 K shells have been produced and parts are now under fabrication. Testing of the first LF detector array will continue, as well as development and fabrication of the HF band array. We expect first light on FYST in 2026. 

\acknowledgments % equivalent to \section*{ACKNOWLEDGMENTS}       
 
The construction of EoR-Spec is supported by NSF grant AST-2009767. The CCAT project, FYST and Prime-Cam instrument have been supported by generous contributions from the Fred M. Young, Jr. Charitable Trust, Cornell University, and the Canada Foundation for Innovation and the Provinces of Ontario, Alberta, and British Columbia. The construction of the FYST telescope was supported by the Gro{\ss}ger{\"a}te-Programm of the German Science Foundation (Deutsche Forschungsgemeinschaft, DFG) under grant INST 216/733-1 FUGG, as well as funding from Universit{\"a}t zu K{\"o}ln, Universit{\"a}t Bonn and the Max Planck Institut f{\"u}r Astrophysik, Garching. G.S., T.N., and R.F. acknowledge support in part by NSF grant AST-1910107. 

% References
\bibliography{report} % bibliography data in report.bib
\bibliographystyle{spiebib} % makes bibtex use spiebib.bst

\end{document}